\input harvmac
\parskip 7pt plus 1pt
\noblackbox

\input epsf

\def\la{\mathrel{\mathpalette\fun <}}
\def\ga{\mathrel{\mathpalette\fun >}}
\def\fun#1#2{\lower3.6pt\vbox{\baselineskip0pt\lineskip.9pt
  \ialign{$\mathsurround=0pt#1\hfil##\hfil$\crcr#2\crcr\sim\crcr}}}
\relax


\lref\kinneyriotto{W.~H.~Kinney, E.~W.~Kolb, A.~Melchiorri and A.~Riotto,
``WMAPping inflationary physics,''
arXiv:hep-ph/0305130.}

\lref\GD{O.~DeWolfe and S.~B.~Giddings,
``Scales and hierarchies in warped compactifications and brane worlds,''
Phys.\ Rev.\ D {\bf 67}, 066008 (2003)
[arXiv:hep-th/0208123].}

\lref\listmyers{R.~C.~Myers,
``Dielectric-branes,''
JHEP {\bf 9912}, 022 (1999)
[arXiv:hep-th/9910053]; J.~Polchinski and M.~J.~Strassler,
``The string dual of a confining four-dimensional gauge theory,''
[arXiv:hep-th/0003136];
C.~Bachas, M.~R.~Douglas and C.~Schweigert,
``Flux stabilization of D-branes,''
JHEP {\bf 0005}, 048 (2000)
[arXiv:hep-th/0003037]; C.~P.~Herzog and I.~R.~Klebanov,
Phys.\ Lett.\ B {\bf 526}, 388 (2002)
[arXiv:hep-th/0111078].} 

\lref\KS{I. Klebanov and M.J. Strassler, ``Supergravity and a
Confining Gauge Theory:  Duality Cascades and $\chi$SB Resolution
of Naked Singularities,'' JHEP {\bf 0008} (2000) 052,
[hep-th/0007191].}

\lref\GKP{S. Giddings, S. Kachru and J. Polchinski, ``Hierarchies
from Fluxes in String Compactifications,'' Phys. Rev. {\bf D66}
(2002) 106006, [hep-th/0105097].}

\lref\nongaussiancurvaton{D.~H.~Lyth, C.~Ungarelli and D.~Wands,
``The primordial density perturbation in the curvaton scenario,''
Phys.\ Rev.\ D {\bf 67}, 023503 (2003)
[arXiv:astro-ph/0208055];
N.~Bartolo, S.~Matarrese and A.~Riotto,
``On non-Gaussianity in the curvaton scenario,''
arXiv:hep-ph/0309033;N.~Bartolo, S.~Matarrese and A.~Riotto,
``Evolution of second-order cosmological perturbations and non-Gaussianity,''
arXiv:astro-ph/0309692.}

\lref\KV{S. Kachru, J. Pearson and H. Verlinde, ``Brane/Flux
Annihilation and the String Dual of a Nonsupersymmetric Field
Theory,'' JHEP {\bf 0206} (2002) 021, [hep-th/0112197].}

\lref\buchel{Buchel and R.~Roiban,
``Inflation in warped geometries,''
[arXiv:hep-th/0311154].}

\lref\deSitter{S.~Kachru, R.~Kallosh, A.~Linde and S.~P.~Trivedi,
``De Sitter vacua in string theory,''
Phys.\ Rev.\ D {\bf 68}, 046005 (2003)
[arXiv:hep-th/0301240].}

\lref\lrreview{D.~H.~Lyth and A.~Riotto,
``Particle physics models of inflation and the cosmological density
perturbation,''
Phys.\ Rept.\  {\bf 314}, 1 (1999)
[arXiv:hep-ph/9807278].}

\lref\guth{A.~H.~Guth,
``The Inflationary Universe: A Possible Solution To The Horizon
And Flatness Problems,''
 Phys.\ Rev.\ D {\bf 23}, 347 (1981).}

\lref\guthweinberg{A.~H.~Guth and E.~J.~Weinberg,
``Could The Universe Have Recovered From A Slow First Order Phase
Transition?,''
Nucl.\ Phys.\ B {\bf 212}, 321 (1983).}

\lref\axion{T.~Banks and M.~Dine,
``The cosmology of string theoretic axions,''
Nucl.\ Phys.\ B {\bf 505}, 445 (1997)
[arXiv:hep-th/9608197];
K.~Choi and J.~E.~Kim,
``Compactification And Axions In E(8) X E(8)-Prime Superstring Models,''
Phys.\ Lett.\ B {\bf 165}, 71 (1985);
K.~Choi and J.~E.~Kim,
``Harmful Axions In Superstring Models,''
Phys.\ Lett.\ B {\bf 154}, 393 (1985)
[Erratum-ibid.\  {\bf 156B}, 452 (1985)];
K.~Choi,
``Axions and the strong CP problem in M-theory,''
Phys.\ Rev.\ D {\bf 56}, 6588 (1997)
[arXiv:hep-th/9706171];
K.~Choi, E.~J.~Chun and H.~B.~Kim,
``Cosmology of light moduli,''
Phys.\ Rev.\ D {\bf 58}, 046003 (1998)
[arXiv:hep-ph/9801280];
T.~Banks and M.~Dine,
``Phenomenology of strongly coupled heterotic string theory,''
[arXiv:hep-th/9609046].}

\lref\dvalitye{G.R.~Dvali and S.-H.~H.~Tye, ``Brane inflation,''
 Phys.\ Lett. {\bf B450} (1999) 72 [arXiv:hep-ph/9812483].}

\lref\others{C.~P.~Burgess, M.~Majumdar, 
D.~Nolte, F.~Quevedo, G.~Rajesh 
and
R.~J.~Zhang,
``The inflationary brane-antibrane universe,''
JHEP {\bf 0107} (2001) 047 [arXiv:hep-th/0105204]; G.~R.~Dvali, Q.~Shafi
and S.~Solganik,
``D-brane inflation,''
[arXiv:hep-th/0105203];  C.~P.~Burgess,
P.~Martineau, F.~Quevedo, G.~Rajesh and R.~J.~Zhang,
``Brane antibrane inflation in orbifold and orientifold models,''
JHEP {\bf 0203} (2002) 052 [arXiv:hep-th/0111025];  M.~Gomez-Reino and
I.~Zavala,
``Recombination of intersecting D-branes and cosmological inflation,''
JHEP {\bf 0209} (2002) 020 [arXiv:hep-th/0207278];
S.~H.~Alexander,
``Inflation from D - anti-D brane annihilation,''
Phys.\ Rev.\ D {\bf 65} (2002) 023507
[arXiv:hep-th/0105032];
B.~s.~Kyae and Q.~Shafi,
``Branes and inflationary cosmology,''
Phys.\ Lett.\ B {\bf 526} (2002) 379
[arXiv:hep-ph/0111101];
J.~H.~Brodie and D.~A.~Easson,
JCAP {\bf 0312}, 004 (2003)
[arXiv:hep-th/0301138].}

\lref\pngb{K.~Dimopoulos, D.~H.~Lyth, A.~Notari and A.~Riotto,
JHEP {\bf 0307}, 053 (2003)
[arXiv:hep-ph/0304050].}

\lref\KKLMMT{
S.~Kachru, R.~Kallosh, A.~Linde, J.~Maldacena, L.~McAllister and 
S.~P.~Trivedi,
JCAP {\bf 0310}, 013 (2003)
[arXiv:hep-th/0308055].}

\lref\kalloshshift{J.~P.~Hsu, R.~Kallosh and S.~Prokushkin,
``On brane inflation with volume stabilization,''
[arXiv:hep-th/0311077].}

\lref\tyeshift{H.~Firouzjahi and S.~H.~H.~Tye,
``Closer towards inflation in string theory,''
[arXiv:hep-th/0312020].}

\lref\rsI{L.~Randall and R.~Sundrum,
``A large mass hierarchy from a small extra dimension,''
Phys.\ Rev.\ Lett.\  {\bf 83}, 3370 (1999)
[arXiv:hep-ph/9905221].}

\lref\curvaton{S.~Mollerach,
``Isocurvature Baryon Perturbations And Inflation,''
Phys.\ Rev.\ D {\bf 42}, 313 (1990); A.~D.~Linde and V.~Mukhanov,
``Nongaussian isocurvature perturbations from inflation,''
Phys.\ Rev.\ D {\bf 56}, 535 (1997);
K.~Enqvist and M.~S.~Sloth,
``Adiabatic CMB perturbations in pre big bang string cosmology,''
Nucl.\ Phys.\ B {\bf 626}, 395 (2002)
[arXiv:hep-ph/0109214];
T.~Moroi and T.~Takahashi,
``Effects of cosmological moduli fields on cosmic microwave background,''
Phys.\ Lett.\ B {\bf 522}, 215 (2001)
[Erratum-ibid.\ B {\bf 539}, 303 (2002)]
[arXiv:hep-ph/0110096];
D.~H.~Lyth and D.~Wands,
``Generating the curvature perturbation without an inflaton,''
Phys.\ Lett.\ B {\bf 524}, 5 (2002)
[arXiv:hep-ph/0110002].}

\lref\decay{G.~Dvali, A.~Gruzinov and M.~Zaldarriaga,
``A new mechanism for generating density perturbations from inflation,''
[arXiv:astro-ph/0303591];
L.~Kofman,
``Probing string theory with modulated cosmological fluctuations,''
[arXiv:astro-ph/0303614].}

\lref\sm{F.~G.~Cascales, M.~P.~G.~del Moral, F.~Quevedo and A.~Uranga,
``Realistic D-brane models on warped throats: Fluxes, hierarchies and
moduli stabilization,''
[arXiv:hep-th/0312051].
}

\lref\tyecosmic{
S.~Sarangi and S.~H.~Tye,
``Cosmic String Production Towards the End of Brane Inflation,''
Phys.\ Lett.\ B {\bf 536} (2002) 185,
hep-th/0204074;
N.~T.~Jones, H.~Stoica and S.~H.~Tye,
``The Production, Spectrum and Evolution of Cosmic Strings in
Brane Inflation,''
Phys.\ Lett.\ B {\bf 563} (2003) 6,
hep-th/0303269;
L.~Pogosian, S.~H.~Tye, I.~Wasserman and M.~Wyman,
``Observational Constraints on Cosmic String Production During Brane
Inflation,''
Phys.\ Rev.\ D {\bf 68} (2003) 023506,
hep-th/0304188.
E.~J.~Copeland, R.~C.~Myers and J.~Polchinski,
``Cosmic F- and D-strings,''
arXiv:hep-th/0312067.
}

\lref\valuewmap{D.~N.~Spergel {\it et al.},
``First Year Wilkinson Microwave Anisotropy Probe (WMAP) Observations:
Astrophys.\ J.\ Suppl.\  {\bf 148}, 175 (2003)
[arXiv:astro-ph/0302209].}
\lref\ng{
E.~Komatsu {\it et al.},
``First Year Wilkinson Microwave Anisotropy Probe (WMAP) Observations: Tests
Astrophys.\ J.\ Suppl.\  {\bf 148}, 119 (2003)
[arXiv:astro-ph/0302223].}

\lref\SW{
N.~Seiberg and E.~Witten,
``The D1/D5 system and singular CFT,''
JHEP {\bf 9904}, 017 (1999),
[arXiv:hep-th/9903224].}

\lref\KKLMMT{
S.~Kachru, R.~Kallosh, A.~Linde, J.~Maldacena, L.~McAllister and S.~P.~Trivedi,
``Towards inflation in string theory,''
JCAP {\bf 0310}, 013 (2003), [arXiv:hep-th/0308055].}

\lref\tonireview{A.~Riotto,
``Inflation and the theory of cosmological perturbations,''
Lectures given at ICTP Summer School on Astroparticle Physics and
Cosmology, Trieste, Italy, 17 Jun - 5 Jul 2002.
Published in *Trieste 2002, Astroparticle physics and cosmology* 317-413 ;
[arXiv:hep-ph/0210162]
 }

\lref\brand{R. Brandenberger, P.-M. Ho and H.-C Kao,
''Large N Cosmology", hep-th/0312288.}

\lref\dvalioldnew{G.~Dvali and S.~Kachru,
``New old inflation,''
[arXiv:hep-th/0309095].
}

\lref\hybrid{A.~D.~Linde, ``Hybrid Inflation,'' Phys.\ Rev.\ D {\bf 49} (1994)
748, [arXiv:astro-ph/9307002].}

\lref\BP{R. Bousso and J. Polchinski, ``Quantization of Four-form
Fluxes and Dynamical Neutralization of the Cosmological
Constant,'' JHEP {\bf 0006} (2000) 006, [hep-th/0004134].}

\lref\MSS{A. Maloney, E. Silverstein and A. Strominger, ``de Sitter
Space in Non-Critical String Theory,'' [hep-th/0205316].}

\hskip 1cm
\vskip 0.2in

\Title{\vbox{\baselineskip12pt \hbox{hep-th/0401004}
\hbox{Bicocca-FT-03-36} }}
{\vbox{\centerline{Old Inflation in String Theory}}}
\vskip .2in
\centerline{Luigi Pilo$^{a}$\footnote{}{luigi.pilo@pd.infn.it,
antonio.riotto@pd.infn.it, Alberto.Zaffaroni@mib.infn.it},
Antonio Riotto$^{a}$ and
Alberto Zaffaroni$^{b}$
}
\vskip .2in
\centerline{$^a$ \it INFN, sezione di Padova, 
Via Marzolo 8, Padova I-35131, Italy}
\vskip .2in
\centerline{$^b$
\it Universit\'a di Milano-Bicocca and INFN, Piazza della Scienza 3, Milano
I-20126, Italy}

\vskip .2in
\baselineskip18pt 
\noindent
We propose a stringy version of the old inflation scenario
which does not require any slow-roll inflaton potential and  is 
based on
a specific example of string compactification with warped metric. 
Our set-up admits the presence of anti-$D$3-branes in the deep infrared region
of the metric 
and  a false vacuum state
with positive  vacuum energy density. The latter is responsible 
for the accelerated period of inflation. The false vacuum 
exists only if the number of anti-$D$3-branes is smaller than
a critical number and the graceful exit from inflation is attained
if a number of 
anti-$D$3-branes
travels from the ultraviolet  towards the infrared region.
The cosmological curvature perturbation is generated through
the curvaton mechanism.

\Date{January 2004}

\eject
\baselineskip20pt 

\newsec{Introduction}

Inflation has become the standard paradigm for explaining the
homogeneity and the isotropy of our observed Universe \lrreview .
At some primordial epoch,  the Universe is trapped in some false vacuum 
and the corresponding vacuum
energy gives rise to an exponential growth of the scale factor. 
During this phase a small, smooth region of size of the order
of the Hubble radius grew so large that it easily encompasses
the comoving volume of the entire presently observed Universe
and one can understand why the observed Universe is  homogeneous and
isotropic to such high accuracy. Guth's original idea \guth\ was that
the end of 
inflation could be initiated by the tunneling of the false vacuum into
the true vacuum during a first-order phase transition. However, 
it was shown that the created bubbles of true vacuum would not
percolate to give rise to the primordial plasma of relativistic degrees
of freedom \guthweinberg . This drawback is solved in slow-roll models of
inflation \lrreview\ where a scalar field, the inflaton, 
slowly rolls down along its potential. The latter
has to be very flat  in order to achieve
a sufficiently long period of inflation which is terminated when 
the slow-roll conditions are violated. The Universe enters subsequently 
into a period of matter-domination during which the energy density is
dominated by the coherent oscillations of the inflaton field around the
bottom of its potential. Finally reheating takes place when the
inflaton decays and its decay products thermalize. 

While building up successful slow-roll inflationary models requires
supersymmetry as a crucial ingredient, the flatness
of the potential is spoiled by supergravity corrections, making it very
difficult to construct a satisfactory model of inflation firmly rooted in 
in modern particle theories having supersymmetry as a
crucial ingredient. The same difficulty is encountered
when dealing with  inflationary scenarios in string
theory.
In brane-world scenarios inspired by string theory a primordial
period of inflation is naturally achieved \dvalitye\
and the role of the inflaton is played by 
the relative brane position in the bulk of the underlying
higher-dimensional theory \others\ . In the exact supersymmetric limit
the brane position is the Goldstone mode associated to the translation
(shift) symmetry and the inflaton potential is flat. 
The 
weak brane-brane interaction breaks the translational  shift symmetry 
only slightly, giving rise to a relatively flat inflaton potential and 
allowing a sufficiently long period of inflation. 
However, the
validity of  brane inflation models in string theory depends on
the ability to
stabilize the compactification volume. This means 
the effective four-dimensional theory has to fix the volume modulus while 
keeping the
potential  for the distance modulus flat. A careful consideration
of the closed string moduli reveals that the superpotential 
stabilization of the  compactification volume typically modifies the 
inflaton potential and  renders it too steep for
inflation \KKLMMT\ . Avoiding this problem requires
some conditions on the superpotential needed
     for inflation  \refs{\kalloshshift,\tyeshift}.

In this paper we propose a stringy version of the old inflation scenario
which does not require any slow-roll inflaton potential and  is 
based on string compatifications with warped metric. 
Warped factors are quite common in string theory compactifications
and arise, for example, in the vicinity of $D$-branes sources.
Similarly, string theory has antisymmetric forms whose fluxes in the 
internal
directions of the compactification typically introduce warping.
In this paper we focus on the Klebanov-Strassler (KS) solution \KS\ ,
compactified as suggested in \GKP\ , which consists
in a non-trivial warped geometry with background
values for some of the antisymmetric
forms of type IIB supergravity. The KS model may be thought as 
the  stringy realization of the Randall-Sundrum model
(RSI)  \rsI\ , where the Infra-Red (IR)
brane has been effectively regularized by an IR geometry, and has been
recently used to embed   the Standard Model on anti-$D$3-branes \sm\ .

From the inflationary point of view, the basic property of the set-up
considered in this paper is that it admits in the deep IR region
of the metric the presence of $p$ anti-$D$3-branes. These anti-branes
generate a positive vacuum energy density as in \deSitter .
We consider a
specific scenario studied in \KV\ where the anti-branes form a
metastable bound state. For sufficiently
small values of $p$, the system sits indeed on a false vacuum state
with positive  vacuum energy density. The latter is responsible 
for the accelerated period of inflation. In terms of the four-dimensional
effective description, the inflaton may be identified with
a four-dimensional scalar field parameterizing the angular
position $\psi$ of the anti-$D$3-branes in the internal directions.
The curvature of the potential
around the minimum is much larger than $H_*^2$, being $H_*$
the value of the Hubble rate during inflation, and slow-roll conditions
are violated. Under these circumstances, inflation would last for ever.
The key point is that, as shown in \KV\ , 
the false vacuum for the potential $V(\psi)$ 
exists only if the number of anti-$D$3-branes is smaller than
a critical number. If a sufficient number of anti-$D$3-branes
travels from the Ultra-Violet (UV)  towards the IR region,
thus increasing the value of $p$, inflation
stops as soon as $p$ becomes larger  than the critical value. At this point,
the curvature around the false vacuum becomes negative, the system
rolls down the supersymmetric vacuum and the graceful exit from inflation 
is attained. 

The nice feature about this scenario is that it may be
considered intrinsically of stringy nature. Indeed, the false vacuum energy
may assume only discrete values. Furthermore, 
even though the
dynamics of each  wandering
anti-$D$3-brane may be described in terms of the effective
four-dimensional field theory by means of a scalar field parameterizing
the distance between the wandering
anti-$D$3-brane and the stack of $p$ 
anti-$D$3-branes in the IR, the corresponding effective 
inflationary model would contain a large number of such scalars
whose origin  might be considered obscure if observed 
from a purely 
four-dimensional observer \brand\ . In this paper we
consider a specific set-up that has been already analyzed in literature
\KV\ , however one could envisage other stringy frameworks sharing
the same properties described here.

The last ingredient we have to account for in order to render our 
inflationary scenario attractive is to explain the origin of cosmological
perturbations. It is now clear that structure in the Universe
comes primarily from an almost scale-invariant superhorizon curvature
perturbation. This perturbation originates presumably from the vacuum 
fluctuation,
 during the almost-exponential inflation,  of some field with mass
much less than the Hubble parameter $H_*$. Indeed, every
such field  acquires a nearly scale-invariant classical perturbation.
In our scenario, the inflaton field mass is not light compare to $H_*$ and
its fluctuations are therefore highly suppressed. 
However, its has been recently proposed that the field
responsible for the observed cosmological perturbations
is  some
 `curvaton' 'field different from the inflaton \curvaton .
During inflation, the curvaton energy density is 
negligible and isocurvature perturbations with
a flat spectrum are produced in the curvaton field. After the end
of inflation, 
the curvaton field oscillates during some radiation-dominated era,
causing its energy density to grow and 
thereby converting the initial isocurvature into curvature 
perturbation. 
This scenario liberates the inflaton from the responsibility
of generating the cosmological curvature perturbation
and therefore  avoids slow-roll conditions. We will show that
in our stringy version of the old inflation it is possible
to find scalar fields which have all the necessary properties to
play the role of the curvaton.

The paper is organized as follows. In \S2 we describe our stringy set-up
deferring the technicalities to the Appendices. In \S3 we describe the 
properties of the inflationary stage and the production of the
cosmological perturbations. Finally, in \S4 we draw our conclusions.
The Appendices provide some of the details about the set-up discussed
in \S2.

\newsec{The set-up} 
We consider a specific example of string compactification with
warped metric that has all features for realizing old inflation
in string theory. 
Here we will give
a summary of the properties of the model, referring to the
Appendices for a detailed discussion of the compactification and
for a derivation of the various formulae.

We are interested in a string compactification where we can insert
branes and antibranes at specific locations in the internal directions.
An explicit solution where one can actually study the dynamics of the
inserted branes is the Klebanov-Strassler (KS) solution \KS, 
compactified as suggested in \GKP . 
The KS solution consists in a non-trivial warped geometry with background 
values for some of the antisymmetric
forms of type IIB supergravity. In particular,
the RR four form $C_{(4)}$ and the (NS-NS and R-R) two forms 
$B_{(2)}, C_{(2)}$ are turned on. There are indeed
$N$ units of flux for  $C_{(4)}$ and $M$ units of flux for $C_{(2)}$ 
along some cycles of the internal geometry. 
 The solution is non
compact and one can choose a radial coordinate in the internal
directions that plays the familiar role of the fifth dimension in
the AdS/CFT correspondence  and in the Randall-Sundrum (RS) models.
The KS solution can be compactified by gluing at a certain radial cut-off a 
compact Calabi-Yau manifold that solves
the supergravity equations of motion. In the compact model, one also adds
 an extra flux $K$ for $B_{(2)}$ along one of the Calabi-Yau cycles \GKP.
Summarizing, the solution is specified by the strings couplings $\alpha^\prime$ and
$g_s$
and by three integer fluxes $N$, $M$ and $K$. 
These numbers are constrained by the tadpole cancellation condition
(charge conservation). 
A detailed discussion of these
constraints can be found in \GKP\ and it is reviewed in the Appendices.
The internal manifold typically has other non-trivial cycles. If needed,
extra parameters can be introduced by turning on fluxes on these cycles.

The KS solution was originally 
found in the the contest of the AdS/CFT correspondence
as the supergravity dual of a confining $N=1$ gauge theory. It represents
the near horizon geometry of a stack of three-branes in a singular 
geometry in type IIB string theory. As familiar in the AdS/CFT correspondence,
by looking at the near horizon geometry of a system of branes, one
obtains a dual description in terms of a supergravity theory.
Notice that in the KS supergravity solution there is no explicit source
for the branes: the there-branes sources have been replaced 
by fluxes of the antisymmetric forms along the 
non trivial cycles of the internal geometry. The reason for this replacement
is that $D$-branes in type II strings are charged under the RR-forms.
The fluxes in the KS solution recall that the background was originally
made with (physical and fractional) three-branes 
charged under $C_{(4)}$ and $C_{(2)}$. However one must remember 
that the correct relation
between branes and fluxes is through a string duality 
(the AdS/CFT correspondence).

To our purposes, the KS model has two important properties.
The first one is that there is a throat 
region where the metric is of the form   
\eqn\warped{
ds^2=h^{-1/2}(r)dx_\mu dx^\mu + h^{1/2}(r)(dr^2+r^2ds_{(5)})}
for $r<r_{UV}$.
Here we have chosen a radial coordinate in the compact directions and
we have indicated with $ds_{(5)}$ the angular part of the internal metric.
For $r>r_{UV}$, \warped\  is glued with a metric
that compactifies the coordinates $r$ and contains most of the
details of the actual string compactification. 
The important point here is that the warp factor $h(r)$
is never vanishing and has a minimal value $h(r_0)$. 
Roughly speaking,
the model is a stringy realization of the first Randall-Sundrum model (RSI),
 where the IR
brane has been effectively regularized by an IR geometry.  
In  first approximation, for $r$ sufficiently large, the reader would
not make a great mistake in thinking to the RSI model with
a Planck brane at $r_{UV}$ and an IR brane at $r_0$ (see Figure 1).  To avoid
confusions, it is important to stress that, in our coordinates, large $r$
corresponds to the UV region and small $r$ to the IR (in the RS literature
one usually considers a fifth coordinate $z$ related to $r$ by $r=e^{-z}$).
\midinsert
\centerline{\epsfxsize=5in\epsfbox{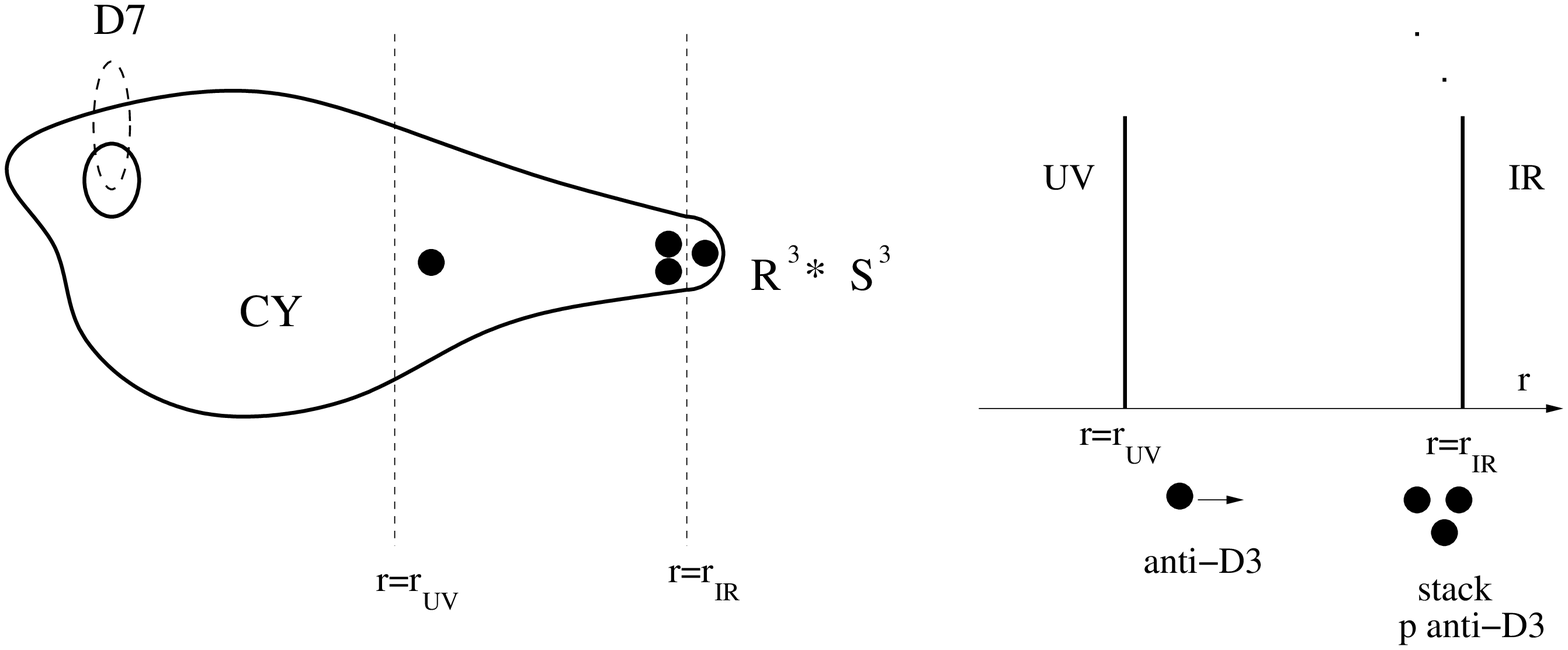}}
Figure 1. The CY compactification with a throat
and its corresponding interpretation in terms of a simplified
RSI model. Recall that small $r$ means IR. 
$D$7 branes that might serve for generating non-perturbative
superpotentials are naturally present in the UV region.
\endinsert

In particular,
for $r\gg r_0$ (but $r<r_{UV}$) the warp factor is approximately
$h(r)=R^4/r^4$ and the metric for the five coordinates $(x_\mu,r)$ 
is Anti-de-Sitter as in the RSI model. In particular, we will
write $\min  h(r)=R^4/r_0^4$ in the following. The precise relation
between parameters identifies \GKP\ 
\eqn\relation{\eqalign{N&=MK, \cr R^4&={27\over 4}\pi g_s N\alpha^{\prime 2}, \cr
 {r_0\over R}&\sim e^{-{2\pi K\over 3g_s M}},\cr
M_{p}^2&={2{\cal V}\over (2\pi)^7 \alpha^{\prime 4}g_s^2},\cr
}}
where ${\cal V}$ is the internal volume and $M_p$ is the four-dimensional
Planck mass. Here ${\cal V}$ is a modulus of the solution even though
the dependence of the warp factor on the volume can be subtle \GKP\ .

The second property deals with the IR region of the metric. We are interested
in putting $p$ anti-$D$3-branes in the deep IR. They will break supersymmetry
and provide the positive vacuum energy necessary for inflation. 
While the RSI model is not predictive about the fate of the anti-branes
in the IR, in the KS model we can analyze their dynamics. It was observed
in \KV\  that the p anti-$D$3-branes form an unstable bound state.
If the background were really made of branes, $p$  anti-$D$3-branes would
annihilate with the existing $D$3 branes. In the actual background, one should 
think that  the anti-$D$3-branes finally annihilate by transforming into pure
flux for $C_{(4)}$, under which they are negatively charged.
The fate of the anti-$D$3-brane can be studied using string theory. The
mechanism \KV , which is reviewed in Appendix III, is roughly as
follows. It is energetically favorable for the anti $D$3-branes
to expand in a spherical shell in the internal directions (see Figure 2).  
The spherical shell is located in the
IR at the point $r_0$ of minimal warp factor. 
The IR geometry is, in a good approximation, $R^7\times S^3$ and 
the branes distribution wraps a two sphere inside $S^3$. 
\midinsert
\centerline{\epsfxsize=3in\epsfbox{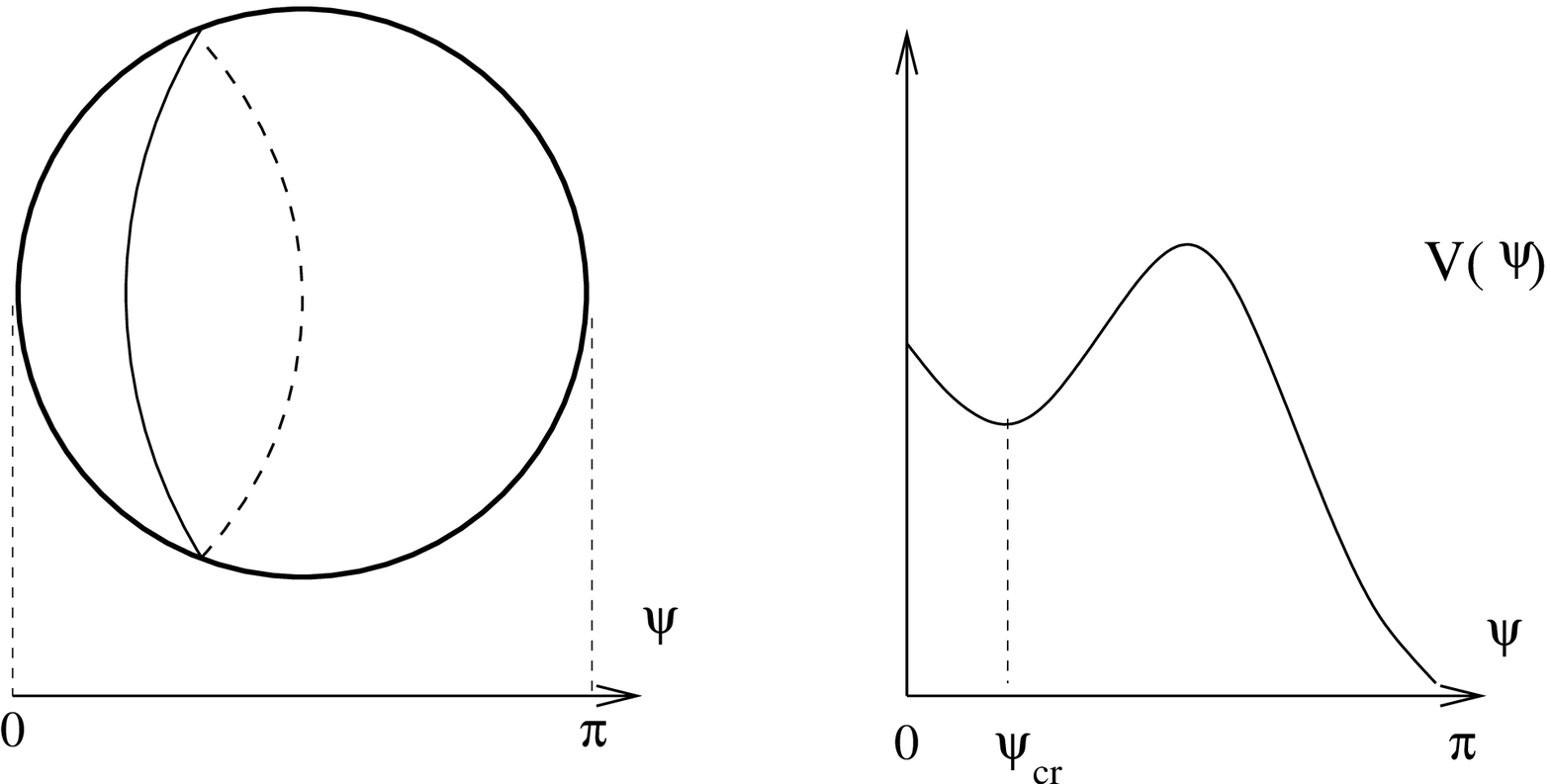}}
Figure 2. The expanded anti-$D$3-branes and the form of the potential
for small values of $p/M$.
\endinsert

The set of all
$S^2\subset S^3$ is parameterized by an angle $\psi\in [0,\pi]$,
where $\psi=0$ corresponds to the North pole and $\psi=\pi$ the South
pole. The angular position $\psi$ of the branes appears as a scalar
in the world-volume action.  As shown in \KV\  the branes feel
a potential in $\psi$. The dynamics of the scalar field $\psi$ can be
summarized by the Lagrangian
\eqn\lagp{L(\psi )=\int d^4 x\sqrt{g}\left \{ M_{p}^2 {\cal R}-
T_3{r_0^4\over R^4}\left [ M
\left( V_2(\psi)\sqrt{1-\alpha^\prime{R^2\over  r_0^2}(\partial\psi)^2}
-{1\over 2\pi}(2\psi-\sin 2\psi )\right ) +p \right]\right \},} 
where $T_3={1\over (2\pi)^3 g_s(\alpha^\prime )^2}$ 
is the tension of the anti-$D$3-branes and 
\eqn\pot{V_2(\psi)={1\over \pi}\sqrt{b_0^2\sin^4\psi+\left({\pi p\over M}
-\psi+{\sin 2\psi \over 2}\right)^2},} with $b_0\equiv 0.9$.

For sufficiently small $p/M$, the potential 
\eqn\pottot{V(\psi)=MT_3({r_0\over R})^4[V_2(\psi)-{1\over 2\pi}(2\psi-\sin 2\psi )+{p\over M}]}
 has the form pictured in 
Figure 2. The original configuration of $p$ anti-$D$3-branes can be identified
with a (vanishing) spherical shell at $\psi=0$. The total energy of the
configuration is 
\eqn\antiDenergy{
V(\psi_{cr})\equiv V_0=2pT_3 \left({r_0\over R}\right)^4,}
where 
$({r_0\over R})^4$ is due to the red-shift caused by the warped metric
and the factor of two is determined by an interaction with the background
fluxes explained in \KV. As shown in Figure 2, it is energetically 
favorable for the branes to expand until $\psi$ reaches the local minimum
at $\psi_{cr}$. The configuration is only metastable; 
the true minimum is at  $\psi=\pi$ where
the shell is collapsed to a point and the energy of
the system vanishes. This means that the anti-branes have disappeared into
fluxes: the final state is supersymmetric and of the same form of the
KS solution with a small change in the fluxes: $M\rightarrow M-p$  and 
in $K\rightarrow K-1$ \KV .

Consider an initial configuration where the bound state of anti-branes
is in the false vacuum $\psi_{cr}$. This provides a vacuum energy
$V(\psi_{cr})$ that causes inflation. For small values of $p/M$,
the critical value of $\psi\sim p/M$ is near zero and the vacuum energy
of the configuration is approximatively given by \antiDenergy .
The mass squared of the fluctuation $\psi$ around the false vacuum
can be computed using
the Lagrangian \lagp\ and reads approximatively 
$m_{\psi}^2\sim (1/\alpha^\prime )(r_0/R)^2$. With a reasonable choice of parameters,
we can easily get  $m_{\psi}^2\gg H_*^2=V(\psi_{cr})/M_{p}^2$, where 
\eqn\hubble{H_*^2={V_0\over 3 M_p^2}\simeq 2p\left({r_0\over R}\right)^4
{T_3\over 3\,M_p^2},}
is the
Hubble rate squared during inflation. This means that the 
field $\psi$ providing the energy density dominating during the
inflationary stage is well fixed at the false ground state.  
The false vacuum can decay to the real
vacuum at $\psi=\pi$ by a tunneling effect but the necessary time,
computed in \KV, is exponentially large. Without any interference
from outside, inflation will last almost indefinitely. 

\subsec{Anti-$D$3-branes in the throat}

Inflation may stop  if extra anti-$D$3-branes are sent in and increase
the value of $p$.
The crucial point is that there is a maximal value $p_{cr}$ 
of $p/M$ for which the potential $V(\psi)$ has a false vacuum. 
For $p>p_{cr}$, the potential is a monotonic decreasing function of $\psi$
(see Figure 3). 

If we send in a sufficient number of anti-$D$3-branes
and $p$ reaches the critical value the false vacuum disappears and
$\psi$ starts rolling down to the real vacuum at $\psi=0$ finishing
the inflationary period. 

\midinsert
\centerline{\epsfxsize=2in\epsfbox{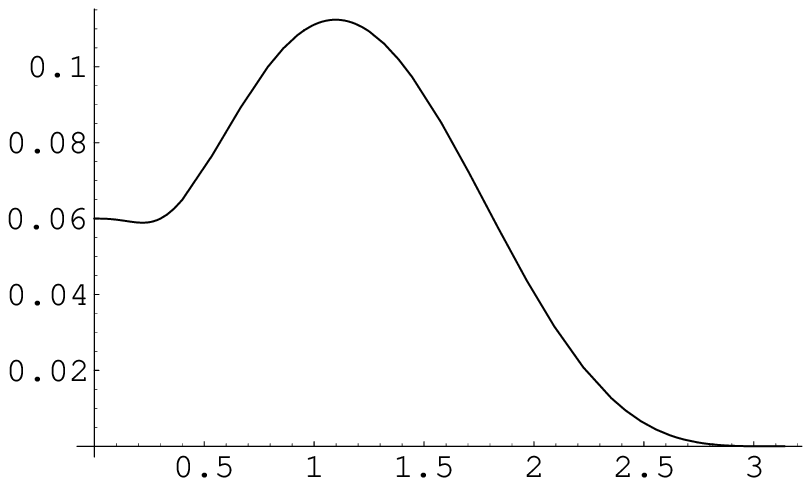}\epsfxsize=2in\epsfbox{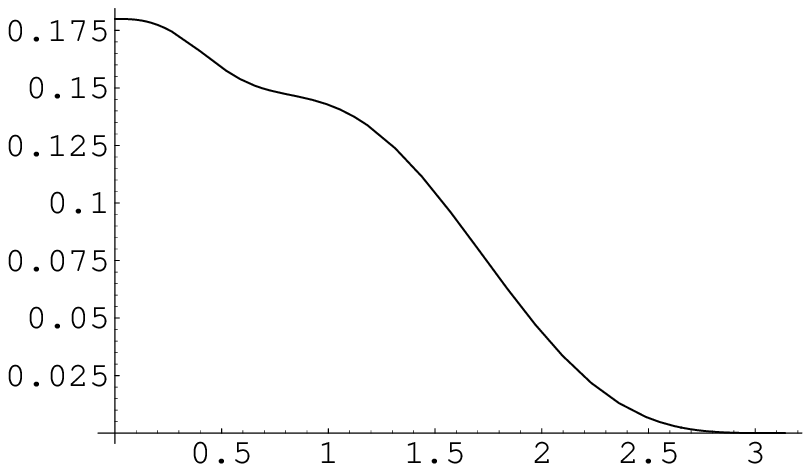}}
Figure 3. The function 
$V_2(\psi)-{1\over 2\pi}(2\psi-\sin 2\psi )+{p\over M}$, equal
to the potential $V(\psi)$ up to an overall scale,
for $p/M=0.03$, where there is a false
vacuum, and for $p/M=0.09$ where the potential is monotonic.
\endinsert

We suppose that the extra anti-branes were originally present
in the compactification at $r>r_{UV}$ and that they left
the UV region of the compactification
for dynamical reasons. 
Their initial energy
at $r=r_{UV}$ is obtained by multiplying the brane tension
with the red-shift factor. For a single anti-brane entering the throat,
the initial energy is $T_3 ({r_{UV}\over R})^4$.
We require that the anti-$D$3-brane is
 a small perturbation of the system
and the vacuum energy is still 
determined by the IR stack of branes. To this purpose,
we must require
\eqn\Denergy{
 T_3 \left({r_{UV}\over R}\right)^4\la  2p\,T_3\, \left({r_0\over 
R}\right)^4.}
To satisfy this condition, we need to consider a mild warping in the throat
or large values of the fluxes. Both requirements can be obtained by varying
the integers $M,K$ (and the internal manifold) 
while keeping order unity values for
the fundamental parameter $g_s$ and $\alpha^\prime\gg M_p^{-2}$,
$T_3\la 1/\alpha^{\prime 2}\ll M_p^4$. For 
instance,
a judicious choice would be  $K\sim M\gg p\ga (R/r_0)^4$ and therefore
$N\sim M^2$. If so, $H^2_*\sim T_3/M_p^2$ and $m^2_\psi /H^2_* \gg 1$
when $\alpha^\prime M_p^2\gg \sqrt{p}/(8\pi^3 g_s)$.  

Once in the throat, the anti-$D$3-brane feels a force toward the IR that can
be estimated as follows. In flat space, there is no-force
between anti-$D$3-branes since they are mutually BPS. In the curved background,
however, there is a force due to gravity and the RR forms. The effective
Lagrangian for the radial position of the brane is computed in Appendix II
 and reads
\eqn\lagr{L(r)=-T_3\int d^4x \sqrt{g}\left ( {1\over 2} 
g^{\mu\nu}(\partial_\mu r)(\partial_\nu r) +{{\cal R}\over 12}r^2 - 
2 h^{-1}(r)\right ) .}
In this equation the potential is twice the red-shift factor, since the 
contribution of the RR forms is equal to that of gravity.
The computation leading to equation \lagr\ is 
similar to that performed in \KKLMMT, where a $D$3-brane 
was moving in the throat. For example, there is the same coupling 
to the Ricci scalar. A crucial difference
with \KKLMMT\ is that, in their case, the potential for a $D$3-brane
was flat and slowly varying. An anti-$D$3-brane has instead a large potential
$V\sim r^4/R^4$
and is rapidly attracted to the IR. 
The Lagrangian \lagr\ is valid for $r\gg r_0$.
Once the anti-brane reaches $r=r_0$, one should analyze the interaction
between anti-$D$3-branes more closely. In first approximation, the net effect
of sending in a single anti-$D$3-brane is to shift $p\rightarrow
p+1$.

\subsec{Volume stabilization}

We suppose that all the moduli of the 
compactification have been stabilized. Backgrounds like the
KS one are particularly appealing because 
all the complex structure moduli of the internal manifold are stabilized
 by  the fluxes \GKP\ . To this purpose, we can also turn on extra fluxes
along the other cycles of the internal manifold.
One K\"ahler modulus, the internal volume, however, is left
massless. This is a typical problem in all string compactifications.
One can imagine, as in \deSitter , that a non-perturbative superpotential is 
generated for the volume. Non-perturbative potentials can be generated,
for example, by the gaugino condensation in gauge groups arising from 
branes in the UV region (see Figure 1) \deSitter .
These non-perturbative effects arising from the 
the UV will not affect the dynamics of the stack of anti-branes
that are located in the IR region. Moreover, $D$7 branes are naturally
present in F-theory compactifications that could serve as a compactification
of the KS solution \GKP . Each strongly coupled gauge factor
 $U(N_c)$ would produce a superpotential for the volume of the form
\eqn\vol{W\sim e^{-2\pi  \rho/N_c},} 
where $\rho={R_{CY}^4\over \alpha^{'2}g_s} +i \sigma$,
$R_{CY}$ being the radius of the internal manifold, 
is a chiral superfield whose real part is related to the internal volume 
and whose imaginary part is an axion-like field. Formula \vol\ follows
from the fact that a $D$7-branes wrapped on four internal directions has
a gauge coupling $1/g_{YM}^2\sim R_{CY}^4/g_s$. We can even suppose
that multiple sets of $D$7 branes undergo gaugino condensation.
In this case we can easily get racetrack potentials consisting of multiple
exponentials where the volume is stabilized with a large mass
while the axion is much lighter, as we will discuss  in the next 
Section. Under these circumstances, the axion will play the role of the
curvaton. 

Notice that both the wandering and the IR anti-D3-branes generates
an extra contribution to the potential for the volume. For example,
the IR stack of branes gives a contribution 
$\sim {1\over (\rho+\bar \rho)^2}$ \refs{\deSitter,\KKLMMT} 
\footnote{$^1$}{This behavior can be 
understood by a Weyl rescaling $g\rightarrow g/{\cal V}$ (in order to decouple
metric and volume fluctuations \GKP ) and by including the warp factor
dependence on the volume \KKLMMT\ .}. Similarly to what supposed
in \KKLMMT\ , we will  assume 
that the scales in the superpotentials are such that the volume 
stabilization is not affected   by the contributions present
during inflation so that the
volume is frozen to its minimum. 
 
As pointed out in \KKLMMT\ , there is the
extra problem that the stabilization of the volume induces a mass
term for the world-volume brane scalar fields of the order of the
Hubble constant. An
explicit coupling to the Ricci tensor has been included in the Lagrangian
\lagr\ for the scalar $r$; it does not affect our arguments as we will 
discuss in the next section.

\newsec{Inflation and the cosmological perturbations}

As we have pointed out in the previous section, the primordial
stage of inflation is driven by the vacuum energy density \antiDenergy\     
stored in the false vacuum provided by a set of $p$ anti-$D$3-branes sitting
in the deep IR region of the metric. 
In terms of the four-dimensional
effective description, the inflaton is identified with
a four-dimensional scalar field parameterizing the angular
position $\psi$ of the anti-$D$3-branes in the internal directions.
The curvature of the potential
around the minimum is much larger than the Hubble rate \hubble\
during inflation. Since slow-roll conditions are not attained,
the curvature perturbation associated to the
quantum fluctuations of the inflaton field $\psi$ is heavily
suppressed. Its amplitude goes like $e^{-m^2_{\psi}/H_*^2}$, with 
$m_\psi\gg H_*$, and the spectrum in momentum space is highly tilted towards
the blue \tonireview\ . 

In our set-up inflation is stopped by whatever mechanism increases
the number of $p$ anti-$D$3-branes beyond some critical value. Indeed, as we
have explained in the previous section, the false vacuum for the potential  
exists only if the number of anti-$D$3-branes is smaller than
a critical number  $p_{cr}$. If the number $p$ changes by an amount
$\Delta p$ (typically a fraction of $M$) becoming  larger  than a critical value, 
the curvature around the false vacuum becomes negative, the system
rolls down the supersymmetric vacuum and the graceful exit from inflation 
is attained. This mechanism is different from what
happens in the four-dimensional hybrid model of inflation \hybrid\  where
inflation is ended by a water-fall transition triggered by the same scalar
field responsible for the cosmological perturbations and is more
reminiscent of the recently proposed idea of old new inflation 
described in Ref. \dvalioldnew\ .

We may envisage therefore the following situation. Suppose that a number of
anti-$D$3-brane is  left in 
the UV region of the compactification
for dynamical reasons.
Once in the throat, each anti-$D$3-brane feels an attractive
force toward the IR proportional to $(r/R)^4$, where
$r$ stands for the modulus parameterizing the distance between each 
anti-$D$3-brane and the IR region. The dynamics of such a modulus
is described in terms of the Lagrangian \lagr\ . Notice, in particular,
that the canonically normalized 
field $\phi=\sqrt{T_3}\,r$ during the inflationary stage receives a contribution
to its mass squared proportional to $H_*^2$,
\eqn\deltam{\Delta m_\phi^2=-{{\cal R}\over 6}=2 H_*^2,}
where we have made of use of the fact that during a de Sitter
phase ${\cal R}=-12 H_*^2$. This contribution spoils the
flatness of the potential in slow-roll stringy models of 
inflation where the inflaton field is identified with the
inter-distance between branes \KKLMMT . In our case, however,
such a contribution is not dangerous and its only effect is to 
suppress  the quantum fluctuations of the field $\phi$. 

Once an  anti-$D$3-brane appears in the UV region, it rapidly flows towards
the IR region under the action of a quartic potential $\lambda\phi^4$, where
$\lambda\sim (T_3 R)^{-1}$, 
and it starts
oscillating around the value $\phi_0=\sqrt{T_3}\,r_0$ under the action of the quadratic
potential $\sim \Delta m_\phi^2\phi^2$. Since the Universe is 
in a de Sitter phase, the amplitude of the oscillations decreases as
\eqn\dec{\phi=\phi_i\,e^{-{3\over 2}({\cal N}-{\cal N}_i)},}
where ${\cal N}={\rm ln} (a/a_i)$ is the number of $e$-foldings and the subscript
$i$ denotes some initial condition. Once the energy density stored in the
oscillations becomes smaller than $\sim 1/\alpha^{\prime 2}$ 
the anti-$D$3-brane
stops its motion and gets glued with the $p$ anti-$D$3-branes in the IR. At this
stage, their number is increased by one unity, going from $p$ to $p+1$. Once
the number of anti-$D$3-brane becomes equal to $p_{cr}$, 
inflation ends since the
system rolls down towards  the supersymmetric vacuum at 
$\psi=\pi$. At this stage
the vacuum energy is released. From the four-dimensional point of view,
this reheating process takes place through the oscillations of the
inflaton field $\psi$ about the minimum of its
potential with mass squared $\sim (1/\alpha^\prime)(r_0/R)^2$. 
From the higher-dimensional point of view, the reheating process 
corresponds to a transition between a metastable
string configuration with fluxes $M,K$ and anti-branes to a stable one.
The anti-branes annihilate releasing energy and changing 
the values of the fluxes, $M\rightarrow M-p, K\rightarrow K-1$.
There is also a complementary description of the reheating process in
terms of the holographic dual. The IR description
of the original system can be given using the dual gauge theory
$SU(2M-p)\otimes SU(M-p)$ (this is the endpoint of the KS cascade \refs{\KS, 
\KV} ). The reheating corresponds to the transition from a metastable 
non-supersymmetric baryonic vacuum to the the supersymmetric one \KV . 
The details of reheating on 
other three or seven branes supporting
our Universe are worth studying in more detail, but this goes
beyond the scope of this paper.

We could try to obtain more realistic IR physics by
introducing other three or seven branes, which might support
our universe. An attempt to embed the Standard Model
of particle interactions in the set-up described in this paper was
recently done in Ref. \sm

\subsec{The number of $e$-foldings}

In our set-up, the total number of $e$-foldings depends  on the initial number of
$p$ anti-$D$3-branes sitting in the deep IR region (the only necessary
condition is $p<p_{cr}$), on the number of wandering  anti-$D$3-branes
in the bulk and, also, on the time interval separating each wandering anti-$D$3-brane
from the next one. 

Due to our ignorance on the
initial state, one can imagine some 
extreme situations. For instance, suppose
that the initial number of
$p$ anti-$D$3-branes sitting in the deep IR region differs from
$p_{cr}$ only by one unity. One wandering anti-$D$3-brane
is therefore enough to stop inflation. Using Eq. \dec\  and taking as
$\phi_i$ the (conservative) value at which the quadratic potential dominates
over the quartic one, the number of $e$-foldings corresponding to
the motion of a single anti-$D$3-brane before capturing is
\eqn\total{{\cal N}\sim{2\over 3}{\rm ln}
\left[{pN\left(r_0/R\right)^4\over M_p^2\alpha^\prime}\right].}
One has to impose   $pN$ to be much larger than the warping factor
$(R/r_0)^4$ in order to get a sizable number of $e$-foldings. This
means that the minimum number $\sim 50$ of $e$-foldings necessary to
explain the homogeneity and isotropy of our observed Universe cannot be
explained in terms of  a single wandering anti-$D$3-brane, but is likely 
to be provided
by the prolonged de Sitter phase preceding the appearance of the
wandering anti-$D$3-brane.

As an alternative, consider the case in which the initial number of
$p$ anti-$D$3-branes sitting in the deep IR region differs from
$p_{cr}$ by several units, say $\sim M$. Under these circumstances
several wandering  anti-$D$3-branes are needed to exit from inflation.
Supposing that the wandering anti-$D$3-branes are well separated in time, the
total number of $e$-foldings between the appearance of the
first anti-$D$3-brane and the end of inflation is given at least as large as
\eqn\totale{{\cal N}\sim {2\over 3} \,M\,
{\rm ln} \left[{pN\left(r_0/R\right)^4\over M_p^2\alpha^\prime}\right].}
We conclude that the last 50 $e$-foldings before the end of inflation
might well  correspond to the period during which $M\sim 50$
anti-$D$3-branes flow into the 
throat.

\subsec{The generation of the cosmological perturbations}

As we have pointed out in several occasions, both the inflaton and the
modulus parameterizing the distance between the
wandering anti-$D$3-branes and the IR region are four-dimensional
degrees of freedom whose mass is much larger than the Hubble rate during
inflation. This implies that their quantum fluctuations are 
not excited during inflation. Fortunately, it has recently become
clear that the curvature adiabatic perturbations 
responsible for the structures of the observed Universe
may well be generated through the quantum fluctuations of some
field other than the inflaton \curvaton\ . The curvaton scenario
relies on the fact that the quantum
fluctuations of any scalar field in a quasi de Sitter epoch
have a flat spectrum as long as the mass of the field is lighter than
the Hubble rate. These fluctuations are of  isocurvature
nature if the energy density of the scalar field
is subdominant. 
The scalar field, dubbed the  curvaton, 
oscillates during some radiation-dominated era,
causing its energy density to grow 
and thereby generating the curvature perturbation.

The requirement that the effective curvaton  mass be much less than the Hubble
parameter during inflation is a severe constraint. In this respect
the situation for the curvaton is the same as that for the inflaton
in the inflaton scenario.
To keep  the effective mass of the  inflaton or curvaton small enough,
it seems natural to invoke supersymmetry and to take
advantage of one of the many flat directions present
in supersymmetric models. However, one has to check 
no  effective mass-squared $\sim H_*^2$ is generated during inflation.

An alternative possibility for keeping the effective mass sufficiently small
is to make the curvaton a pseudo Nambu-Goldstone
boson (PNGB), so that its potential vanishes  in the limit
where  the  corresponding global  symmetry is unbroken.
Then the effective mass-squared of the curvaton 
 vanishes in the
limit of unbroken symmetry and can indeed be kept small by keeping the breaking
sufficiently small. The curvaton as a PNGB has been studied in 
detail in Ref. \pngb\ . 

As we have anticipated in \S2,  non-perturbative superpotentials are
expected to be   
generated for the volume modulus. They can be generated,
for example, by the gaugino condensation in gauge groups arising from 
branes in the UV region and wrapped on four internal directions. This happens
for $D$7 branes which are naturally
present in F-theory compactifications that could serve as a compactification
of the KS solution \GKP . If multiple sets of $D$7 branes undergo gaugino condensation,
one can get racetrack potentials consisting of multiple
exponentials. If we suppose that  the K\"ahler potential does not depend
upon the imaginary part of the volume modulus $\rho$, let us call it 
$\sigma={\rm Im}\,\rho$, and if the non-perturbative superpotential
is of the form
\eqn\spot{W\sim e^{-a\rho}+e^{-b\rho}+\cdots}
where $a<b$ are some positive constants, then the 
axion-like field $\sigma$ receives
a mass which is suppressed by the  
exponential $\sim e^{-(b-a)\,{\rm Re}\,\rho}$ with
respect to the mass of the volume $m_{\cal V}$  \axion\
\eqn\axionmass{m_\sigma^2\sim e^{-(b-a)\,{\rm Re}\,\rho}\,m^2_{\cal V}.}
Because of the exponential suppression, 
the condition $m_\sigma^2\ll H_*^2$ during inflation does not
require any particular fine-tuning.
The axion $\sigma$  plays the role of the
curvaton. Furthermore, since the non-perturbative superpotentials
arise in the UV region, no warping suppression is expected and the
axion scale $f$ will be of the order of $M_p$ 
in the four-dimensional effective theory.
The condition $m_\sigma/f\ll 10^{-2}$ imposed in order to be sure that
inflation lasts enough for the curvaton to be in the quantum regime \pngb\   is
likely to be satisfied.

The requirement that the 
curvaton potential be negligible during inflation corresponds to
\eqn\condition{f m_\sigma \ll M_p H_*.}
Since it  is assumed that the curvaton is light during inflation,
$m_\sigma\ll H_*$,   on super-horizon scales the curvaton has a classical
perturbation with an almost  flat spectrum given by
\eqn\spectrum{\langle\delta\sigma^2\rangle^{1\over 2} = {H_*\over 2\pi}.}
When, after inflation,  $H\sim m_\sigma$, 
 the field 
starts to  oscillate around zero. At this stage the 
 curvaton energy density is
$\rho_\sigma = {1\over 2}m_\sigma^2 \sigma_*^2$ while the total is
$\rho\sim H^2M_p^2$. Here $\sigma_*$ is the value of the curvaton during
inflation.
The fraction of energy stored in the
curvaton is therefore $\sim (\sigma_*/M_p)^2$, which is 
small provided that $\sigma_*\ll M_p$.

After a few Hubble times the oscillation will be sinusoidal
except for the Hubble damping. The  energy density $\rho_\sigma$
will then be proportional to the square of the oscillation amplitude,
and will scale like the inverse of the locally-defined comoving volume
corresponding to matter domination. On the spatially flat slicing, 
corresponding to uniform local expansion, its perturbation has a constant
value
\eqn\dd{{\delta\rho_\sigma\over \rho_\sigma} = 2q \left({\delta\sigma\over\sigma} 
\right)_*.}
The  factor $q$  accounts for the evolution of the field
from the time that $m_\sigma/H$ becomes significant, and will be
  close to 1 provided that  $\sigma_*$ is not too close
to  the maximum value $\pi v$.
The curvature perturbation $\zeta$ is supposed to be negligible when the
curvaton starts to oscillate, growing during some radiation-dominated
era when $\rho_\sigma/\rho\propto a$.
After the curvaton decays $\zeta$  becomes constant. In 
the approximation that the curvaton decays instantly
(and setting $q=1$) it is then given by \curvaton
\eqn\zetaprim{\zeta \simeq {2\gamma\over 3} \left({\delta \sigma\over \sigma} \right)_*,}
where 
\eqn\r{
\gamma\equiv \left. {\rho_\sigma\over\rho} \right|_{D},}
and the subscript $D$ denotes the epoch of decay. The corresponding spectrum
is 
\eqn\a{
{\cal P}_\zeta^{1\over 2}\simeq {2\gamma\over 3}  \left({H_*\over 2\pi \sigma_*}\right).}
It  must
match the observed  value $5\times
10^{-5}$    \valuewmap\    which means that
$H_*/2\pi \sigma_*\simeq 5 \times 10^{-4}/\gamma$.
The current
WMAP bound on non-gaussianity \ng\     requires $\gamma\ga 9\times 10^{-3}$.
In terms of the fundamental parameter of our theory, we get
\eqn\conditionfinal{{p\, T_3\over M_p^4}\left({r_0\over R}\right)^4\sim 
{10^{-6}\over \gamma^2}\la 10^{-2},}
where we have taken $\sigma_*\sim f\sim M_p$.

Before closing this section, we would like to mention  a possible alternative
for the curvaton field. 
During its motion toward the
IR, the anti-$D$3-brane can fluctuate in the internal angular directions.
The scalar fields associated with the angular positions are almost
massless. In particular, being angles, 
they do not get masses from the volume stabilization
mechanism. In the K\"ahler potential
of four-dimensional supergravity the volume modulus $\rho$ always appears in 
the
combination $\rho+\bar\rho - k(\phi_i,\bar\phi_i)$ \GD , 
where $\phi_i$ collectively
denote the position of the branes in the six internal directions,
and $k$ is the K\"ahler potential for the geometry.
This coupling generates a mass for the fluctuations $\phi_i$.
However, at least for large values of $r$,
the geometry has several isometries and
$k(\phi,\bar\phi)$ does not depend on some of the angles. 
For example, for large $r$ the geometry of the KS throat is that of
a cone over the Einstein manifold $T^{1,1}=SU(2)\otimes SU(2)/ U(1)$.
The internal geometry has therefore the isometry $SU(2)\otimes SU(2)\otimes 
U(1)$ that guarantees the independence of $k(\phi,\bar\phi)$ from some
angles. In this way, the 
form of potential for the angular
fluctuations is not affected by the stabilization mechanism and leaves
some angles much lighter than the Hubble rate during inflation and they
may play the role of the curvaton(s).
The isometries do disappear in the IR region, since there the singular
cone over $T^{1,1}$ is made smooth by deforming the tip of the cone.
One then expects that such curvatons becomes massive after inflation 
and  decay during or after reheating. If so, one expects
$\gamma\la 1$ thus enhancing the
non-Gaussian signature \nongaussiancurvaton\ . 

Another relevant prediction of our model is that the Hubble rate
during inflation is not necessarily small. This is different
from what usually assumed in models which make use of the curvaton
mechanism to produce cosmological perturbations where the Hubble rate
is tiny in order to suppress the curvature  perturbations from the
inflaton field. In our set-up the latter are suppressed on
superhorizon scales not by the
smallness of $H_*$, but by the fact that the inflaton field
$\psi$ is not light during inflation. Therefore, a
generic prediction of
our model is that gravitational waves may be produced at an observable
level and close to the present bound corresponding to
$H_*\la 10^{14}$ GeV \kinneyriotto\ .

\newsec{Conclusions}

Brane-world scenarios in string theory offer new ways of obtaining a primordial
period of inflation necessary to account for the homogeneity and isotropy 
of our observed Universe. At the same time, they pose some
challenges. In warped geometries, implementing volume stabilization spoils
the flatness of the inflaton potential in brane-antibrane inflation \KKLMMT\
unless a shift symmetry is preserved 
\refs{\kalloshshift,\tyeshift}.

In this paper we have described a specific 
example of string compatification with warped metric which leads to 
the old inflation scenario and does not require any slow-roll inflaton potential.
Warping is introduced by antisymmetric forms with nonvanishing  fluxes in the 
internal
directions of the compactification.  A stack of anti-$D$3-branes reside
in the deep IR region of the warped metric and, if their number is not larger than
a critical value, supersymmetry is broken and a false vacuum is formed.
From the four-dimensional point of view, the system is described in terms
of a scalar field parameterizing the position of the antibranes along the
internal directions. Such a scalar is well anchored at the false vacuum
with a mass much larger than the Hubble rate during inflation. Slow-roll
conditions are violated. Graceful exit from inflation
is attained  augmenting the number of antibranes in the
IR region by sending antibranes towards the IR from the UV region
of the warped geometry. These single wandering antibranes are inevitably attracted
by the stack of antibranes in the IR and, after a few oscillations, end up
increasing the number of the antibranes in the stack thus stopping inflation. 
Cosmological curvature perturbations are generated through the curvaton
mechanism. We have primarily focused
on the imaginary part of the volume modulus as a curvaton
field,  showing that its mass can be easily 
lighter than the Hubble rate during inflation. The curvaton acts
as a PNGB whose dynamics has been thoroughly studied in \pngb\ .

There are interesting issues which would deserve further and careful
investigation. First of all, we have not studied in this paper
the process of reheating. From the four-dimensional point of view,
reheating  takes place through the oscillations of the
inflaton field $\psi$ about the minimum of its
potential with mass squared $\sim (1/\alpha^\prime)(r_0/R)^2$. 
From the higher-dimensional point of view, reheating corresponds to the
disappearance of the stack of antibranes and the appearance of 
fluxes. The final state is supersymmetric and of the same form of the
KS solution with a small change in the fluxes: $M\rightarrow M-p$  and 
in $K\rightarrow K-1$. Using the holographic duality, we can identify
the final state with an $SU(2M-p)\otimes SU(M-p)$ supersymmetric
gauge theory. We could try to obtain more realistic IR physics by
introducing other three or seven branes, which might support
our universe. An attempt to embed the Standard Model
of particle interactions in the set-up described in this paper was
recently done in Ref. \sm . With a specific model at hand the details
of the transition to the final state and the reheating process could
be studied. 
It would be also interesting to see whether this transition leaves behind
topological defects \tyecosmic\ and which is, eventually, their impact of the
subsequent cosmological evolution. 
We will come back to all these issues in the next future.

\vskip 1cm

\centerline{\bf Acknowledgments}

We would like to thank R. Kallosh and A. Linde for useful discussions.
A.Z. is partially
supported by INFN and MURST under contract 2001-025492, and by 
the European Commission TMR program HPRN-CT-2000-00131.

\appendix {I}{The KS solution}
Warped solutions with a throat of the form \warped\ are common in string
theory. They can be generated by either a stack of branes
or by using solutions with RR fluxes. The two pictures (branes versus fluxes)
are dual to each other in the sense of the AdS/CFT correspondence.
We will consider solutions with fluxes.
One can take many examples out of the AdS/CFT literature. In this context
one usually consider non-compact solutions with a radial coordinate $r$.
To obtain a compact 
model, one must truncate the metric at a certain UV scale $r_{UV}$
and glue a compact manifold  for $r>r_{UV}$.
Varying the internal manifold and the combinations of brane/fluxes, 
one can engineer various supersymmetries. 
 
For example, if we choose $h(r)=R^4/r^4$ and the metric for 
the round five-sphere
for $ds_{(5)}$, we obtain 
\eqn\AdS{
ds^2={r^2\over R^2}dx_\mu dx^\mu + {R^2\over r^2}dr^2 +R^2ds_{S_5},}
the product of $AdS_5\times S^5$. The solution also contains $N$ units of
flux for the RR four-form $C_{(4)}$ along $S^5$.
This choice of warp factor corresponds to a maximally 
supersymmetric solution of
string theory and it is equivalent to the RSII model. The compact manifold  
glued for $r>r_{UV}$ corresponds to an explicit realization of the Plank brane
of the RS scenario. The RSI model can be obtained by truncating the 
metric~\warped\
at $r=r_0$ by the insertion of an IR brane. In contrast to the RSII
model, the warp factor is now bounded above zero and has a minimal
value that has been used to study the hierarchy problem.

The IR brane can be replaced by any regular geometry that has a non-zero 
minimal warp factor. A  regular type IIB solution with background fluxes 
with this property has been found by Klebanov and Strassler. 
In terms of an appropriate radial coordinate $\tau$ for which the IR
corresponds to $\tau=0$, the KS solution has a the form
\eqn\klebsol{ds^2=h^{-1/2}(\tau)dx_\mu dx^\mu + h^{1/2}(\tau)ds_{(6)}^2,}
with \eqn\detail{\eqalign{ 
h(\tau)&= {\rm const}\times \, I(\tau),\cr
I(\tau )&=
\int_\tau^\infty {x\coth x -1\over \sinh^2 x} (\sinh (2x) - 2x)^{1/3}\cr}}
and with a complicated internal metric, which depends on $\tau$ and five
angles. Here $ds_{(6)}$ is the metric for a deformed conifold. 
It is obtained by
taking a cone over a five-dimensional Einstein manifold ($T^{1,1}$) with
the topology of $S^3\times S^2$ and by deforming the tip of the cone
in order to have a smooth manifold. The resulting manifold has a non-trivial
$S^3$ cycle. In addition to the non-trivial metric, 
there are fluxes for the antisymmetric forms of type IIB supergravity.
The solution preserves $N=1$ supersymmetry. 

The non-compact KS solution can be embedded in a a genuine
string compactification as explained in \GKP . The most convenient
way is to consider F-theory solutions that can develop a local conifold
singularity. An explicit example is provided in \GKP . 
In the compact solution, the R-R and NS-NS two-forms
have integer fluxes along the $S^3$ cycle of the conifold (call it A)
and along its Poincar\'e dual B, respectively:
\eqn\fluxes{{1\over (2\pi)^2\alpha^\prime}\int_A F=M,\,\,\,\,\,\,\,\,\,\,
 {1\over (2\pi)^2\alpha^\prime}\int_B H=-K,}
where $F$ and $H$ are the curvatures of $C_{(2)}$ and $B_{(2)}$.
In order to avoid large curvature in the solution that would invalidate 
the supergravity approximation, the integers $M$ and $K$ must be large.
The solution has a minimal warp factor that is given by
$e^{-{2\pi K\over 3g_s M}}$.

We can include wandering $D$3 and anti-$D$3-branes in the compactification.
For this we must ensure that the total $D$3-charge is zero, as required by
Gauss law in the case of a compact manifold. The effective $D$3-charge
gets contribution from  $D$3 and anti-$D$3-branes and from the various 
couplings
of $C_{(4)}$ to the two-forms and to the curvature of the internal manifold.
The resulting constraint is 
\eqn\charge{{\chi\over 24}=N_3-\bar N_3+KM,}
where $N_3,\bar N_3$ are the number of branes and anti-branes and $\chi$
is the Euler characteristic of the manifold used for the F-theory
compactification. In all our examples, $N_3=0$ and the number of anti-branes
p is much smaller than the background flux M. Defining
$N=\chi/24$ the effective $D$3-charge,  we have to
satisfy $N=KM$. 

We will only need the asymptotic behavior of the metric \klebsol\ for 
large and small radial coordinate, where it assumes the form  given
in \warped . For large $\tau$, it is convenient to use the variable 
$r^2\sim e^{2\tau/3}$ and we have
\eqn\largerad{h(r)={R^4\over r^4} \left (1+{\rm const}\, \log {r\over r_{cr}}\right )}
which corresponds to a logarithmic deformation of AdS.
For small $\tau$, $h(\tau)$ approaches a constant and
the internal metric $ds_{(6)}$ is the product 
$R^3\times S^3$. The radius square of $S^3$ (measured in ten dimensional units)
 is of order  $g_sM\alpha^\prime$
and this quantity must be large for the validity of the supergravity 
approximation. 

According to the holographic
interpretation of the RS model, the IR part of the geometry corresponds
to four-dimensional matter fields that are determined by using the
AdS/CFT correspondence: the IR part of the metric in the RSII model
corresponds to a CFT theory, the IR part of the metric of the KS
solution corresponds to a pure SYM theory. In particular, the holographic
dual of the KS solution, for $p=0$ and $N$ multiple of $M$ ($N=MK$), 
is  a $SU(N+M)\otimes SU(N)$ gauge 
theory, which undergoes a series of Seiberg duality leaving a pure 
confining $SU(M)$ SYM theory in the IR \KS . For $p\ne 0$, we can expect that
the anti-D3-branes annihilate $p$ physical branes giving a 
$SU(N+M-p)\otimes SU(N-p)$ gauge theory. As discusses in \refs{\KS,\KV} ,
the Seiberg duality cascade stops at $SU(2M-p)\otimes SU(M-p)$, a gauge
theory with still a moduli space of vacua.
  
\appendix{II}{The wandering anti-$D$3-branes}
An anti-$D$3-brane entering the throat region at $r\gg r_0$ can be considered as
a probe. It will feel a force toward the IR region.
Anti-$D$3-branes are mutually BPS and, therefore, in first approximation,
the contribution to the force 
from the IR stack of p anti-branes can be neglected.
  The wandering anti-brane will feel a potential due to the non-trivial
warp factor and the RR-fields background. 
We also suppose that the stack of IR branes is not modifying the
background and it has the only effect to induce a non-zero vacuum energy.
In a more precise calculation, one should consider a de-Sitter deformed
KS solution \buchel . 

In the probe approximation, the back-reaction on the metric can be
neglected. The fields in the world-volume effective action couple to
the metric and to all background antisymmetric forms. Our background 
for $r\gg r_0$ has the form given in \warped\ and a non-vanishing four-form
\eqn\fourf{C_{(4)}\sim h^{-1}(r)\epsilon_{0123}.} 
We write the effective action for the position of a $D$3 or anti-$D$3-brane
probe, 
\eqn\probe{S=-T_3\int d^4 \sqrt{g_{IND}} +qT_3\int C_{(4)}.}
The first term in this equation is the Born-Infeld action that depends on
the induced metric and the second term is the Wess-Zumino coupling to the
RR fields. 
Here $T_3={1\over (2\pi)^3(\alpha^\prime)^2g_s}$ is the tension of the brane
and q is the charge under $C_{(4)}$ which is +1 for $D$3-branes and -1 for 
anti-$D$3-branes.  
With the given values for the background fields,
\eqn\dbbb{S=-T_3\int d^4x h^{-1}(r)\sqrt{1-h(r)(\partial r)^2} +
qT_3\int d^4 h^{-1}(r).}

We see that the potential energy for a brane is given by two contributions,
one coming from the non-trivial red-shift of the metric and a second one
coming from the RR fields. A $D$3-branes feels no force in this background
\eqn\db{V(r)=-T_3h^{-1}(r)+T_3h^{-1}(r)=0.}
This fact can be understood easily if one invoke the duality that relates
our background with fluxes to systems of $D$3-branes. The KS solution
is dual to an $N=1$ gauge theory with a moduli space of vacua.
In the language of branes, this corresponds to the possibility to 
separate one or more $D$3-branes from the stack with no cost in energy.
The gravitational attraction between branes is compensated by the 
charge repulsion due to the RR fields. The fact that a $D$3-brane is
a BPS object in the background was used in \KKLMMT\ to obtain
an almost flat potential for the moving brane.

On the other hand, the anti-$D$3-branes will feel a potential
\eqn\antidb{V(r)=-T_3h^{-1}(r)-T_3h^{-1}(r)=-2T_3h^{-1}(r).}
In this case, indeed, the charge of the anti-$D$3-branes has changed sign
and the Coulomb and gravitational forces will add.

In the Lagrangian for the anti-$D$3-branes we should also include a 
coupling to the four-dimensional curvature ${\cal R}$. To our purpose, 
we can approximate the metric for large $r$ with an AdS metric.
In a five-dimensional
AdS background this coupling has been computed in \SW\ for large $r$.
The coupling is generated by the Born-Infeld part of the action and
therefore has the same sign for both $D$3 and anti-$D$3-branes. The
result is that of a conformally coupled scalar. 

Including all contributions the effective action
for the anti-$D$3-branes reads (up  to two derivatives)
\eqn\lagrtwo{L(r)=-T_3\int d^4x \sqrt{g}\left ( {1\over 2} 
g^{\mu\nu}(\partial_\mu r)(\partial_\nu r) +{{\cal R}\over 12}r^2 - 
2 h^{-1}(r)\right ) .}

\appendix{III}{The IR anti-branes}
Our purpose in this Section is to explain formula \lagp\
\eqn\lagptwo{L(\psi )=-T_3\int d^4 x\sqrt{g} {r_0^4\over R^4}\left [ M
\left( V_2(\psi)\sqrt{1-{1\over r_0^2}(\partial\psi)^2}
-{1\over 2\pi}(2\psi-\sin 2\psi )\right ) +p \right],}
giving a brief account of how it is obtained. Details can be found in \KV .
In this Section we put $\alpha^\prime =1$.


In the IR the internal metric is the product $R^3\times S^3$.
The total geometry is $R^7\times S^3$ with the radius 
square of $S^3$ being $b_0^2g_sM\alpha^\prime$ ($b_0^2\sim 0.9$). 
As already mentioned, this 
quantity must be large for the validity of the supergravity
approximation.
We can choose the metric for $S^3$ as
\eqn\mets{b_0^2g_sM (d\psi^2+\sin^2\psi d\Omega_{(2)}).}
There are also non-trivial RR and NS-NS form. From the
condition $\int_{S^3} F =4\pi^2 M$, one easily gets the potential
\eqn\cp{C_{(2)}=4\pi M\left(\psi-{\sin 2\psi\over 2}\right)d\Omega_{(2)}}
With a more accurate computation using the equations of motion one can
also determine $H_{(3)}$ and its dual $H_{(7)}=* H_{(3)}$
($ H_{(7)}=dB_{(6)}$) \KV .

The crucial observation for studying the dynamics of the system is
that a NS-brane wrapped on a two sphere in $S^3$ with p units 
of world-volume flux has the same quantum
number of the stack of p anti-branes. This is a standard observation
in string theory (for a partial list of references related to our configuration
 see \listmyers ). Consider indeed a NS-brane wrapping the cycle
specified by the angle $\psi$. It has a world-volume action
\eqn\NSlagr{S=-{\mu_5\over g_s^2}\int d^6x \sqrt{G_{IND}+
g_s(2\pi F_{(2)}-C_{(2)})} +\mu_5\int d^6x B_{(6)}
}
This action can be obtained by S-duality from the Born-Infeld and Wess-Zumino
action for D5 branes. In string units $\alpha^\prime =1$, 
$\mu_5={1\over (2\pi)^5}=
{\mu_3\over 4\pi^2}$ is the tension of a $D$5-brane. $F_{(2)}$ is the $U(1)$
world-volume field of the brane that, for gauge invariance, must always
couple to $C_{(2)}$. Beside the Wess-Zumino term $B_{(6)}$,
corresponding to the unit charge of the NS brane, there is a 
Wess-Zumino coupling $C_{(4)}\wedge (2\pi F_{(2)}-C_{(2)})$.
This term is responsible for the induced anti-$D$3-charge
when we introduce a non-zero flux for $F_{(2)}$
on the two-sphere,
\eqn\fluxf{\int_{S^2} F_{(2)} =2\pi p.}
When the NS brane is at $\psi=0$,
the two-sphere is vanishing and the five-brane becomes effectively a three
brane. This three brane is not tensionless because $F_{(2)}$ enters
explicitly in the Born-Infeld action. By reducing the action \NSlagr\
on the vanishing two-sphere we obtain a three brane with tension
$4p\pi^2 T_5=pT_3$ 
and negative $D$3-charge -p. These are the quantum numbers
of a stack of $p$ anti-$D$3-branes.

The description in terms of a wrapped NS brane it useful when $\psi\ne 0$
and it shows that the stack of branes can lower its energy by expanding
into an NS brane. Formula \lagptwo\ is obtained from \NSlagr\ by
integrating on the two-sphere. 
The potential
\eqn\pottwo{V_2(\psi)={1\over \pi}\sqrt{b_0^2\sin^4\psi+\left({\pi p\over M}
-\psi+{\sin 2\psi \over 2}\right)^2}}
is the contribution of the determinant of the two by two matrix 
$G_{IND}+
g_s(2\pi F_{(2)}-C_{(2)})$ in the directions of $S^2$. The contribution
 $\sim 2\psi-\sin 2\psi $
to the potential comes from the integral of $B_{(6)}$. There is no
contribution from $C_{(4)}$ since the four-form  vanishes in the IR \KS .
Finally the contribution of p units of $D$3 tension to \lagptwo\
comes from the background fluxes via the tadpole cancellation condition 
\KV . Finally, the potential \lagptwo\ is obtained by introducing the 
appropriate warp factor everywhere.   
The total potential in $\psi$ has
a form that depends on p and it is pictured in Figure 3.
The real minimum is at $\psi=\pi$ and it has vanishing energy.


\listrefs

\bye